# Evidence for Undoped Weyl Semimetal Charge Transport in $Y_2Ir_2O_7$

P.G. LaBarre[1], L. Dong[2,3], J. Trinh[1], T. Siegrist[2,3], A. P. Ramirez[1]

[1]*Physics Department, University of California Santa Cruz, Santa Cruz, CA 95064.*
[2]*FAMU-FSU College of Engineering, Tallahassee, FL 32310, USA.*
[3]*National High Magnetic Field Laboratory, Tallahassee, FL 32310, USA.*

**Abstract**

Weyl fermions scattering from a random Coulomb potential are predicted to exhibit resistivity versus temperature $\rho \propto T^{-4}$ in a single particle model. Here we show that, in closed-environment-grown polycrystalline samples of $Y_2Ir_2O_7$, $\rho = \rho_0 T^{-4}$ over four orders of magnitude in $\rho$. While the measured prefactor, $\rho_0$, is obtained from the model using reasonable materials parameters, the $T^{-4}$ behavior extends far beyond the model's range of applicability. In particular, the behavior extends into the low-temperature, high-resistivity region where the Ioffe-Regel parameter, $k_T l \ll 2\pi$. Strong on-site Coulomb correlations, instrumental for predicting a Weyl semimetal state in $Y_2Ir_2O_7$, are the possible origin of such "bad" Weyl semimetal behavior.

The importance of topologically non-trivial materials is well established and is leading to a close examination of their physical properties. One type of topological system is the Weyl semimetal (WSM), which possesses opposite chirality pairs of intersecting linearly dispersing electronic bands. The identification of WSM systems has relied primarily on the observation of linear dispersion or Fermi arcs by angle resolved photo-electron spectroscopy in materials such as TaAs and ZrTe$_5$ [1,2], accompanied by negative longitudinal magnetoresistance, a result of the so-called "chiral anomaly". In zero magnetic field, however, the resistivity versus temperature ($\rho(T)$) of many WSM compounds is metallic ($d\rho/dT > 0$) [3], or is not described by a simple form [2]. These $\rho(T)$ signatures can be viewed as arising from doping, either intrinsic or extrinsic, which places the chemical potential at an energy different from the intersecting nodes [4-6].

Among candidate WSM materials, the iridate pyrochlores, $R_2Ir_2O_7$ (R = rare earth element) have attracted much attention since calculations using the LSDA + U + SO (LSDA = local spin density approximation, U = electronic correlation parameter, SO = spin orbit) method suggest an interplay between WSM and Mott insulating states for intermediate and strong correlations respectively [7,8]. These compounds develop an "all-in-all-out" (AIAO) magnetic structure among Ir local moments which is thought to create time reversal symmetry breaking leading to a WSM state with 24 Weyl cones whose nodes all lie at the same energy [8]. Despite the simplicity of this predicted electronic structure, reported transport measurements show a large variation among different samples, suggesting a proclivity to doping during synthesis [7-15]. Results on single crystals grown in a closed environment, on the other hand, imply that stoichiometric $R_2Ir_2O_7$ can be thought of as undoped, i.e. the chemical potential lies precisely at the Weyl node energy [16].

In the present work, we re-visit charge transport in $Y_2Ir_2O_7$ with samples grown in a closed environment and over a narrow range of uncharged impurity incorporation. The absence of magnetism from Y allows for a simpler interpretation of the resistivity, $\rho(T)$, compared to other $R_2Ir_2O_7$ systems where the R moment can introduce additional magnetic scattering. We find that $\rho(T) = \rho_0 T^{-4}$ over four orders of magnitude in $\rho(T)$ for $T < 50$K, in agreement with a theory of thermally screened charged impurities (TSCI) [4-6]. The magnitude of the prefactor, $\rho_0 = 2.5 \times 10^9 \Omega$cm, is consistent with estimates using reasonable values for the materials parameters. A discrepancy is found, however, between the temperature range over which $\rho(T) \propto T^{-4}$ is observed and the predicted cross over to maximum resistivity, which is well above this range. This discrepancy demonstrates a limitation of the TSCI model for Y$_2$Ir$_2$O$_7$, which we suggest can be removed by including electronic correlations. The stability of the $T^{-4}$ behavior with Bi incorporation supports our claim of near-stoichiometry of the present samples. We also present specific heat ($C(T)$) data and argue that, within the above theoretical framework, the expected $C(T)$ of the Weyl fermions is orders of magnitude smaller than the phonon contribution. The anomalously large low-temperature $C(T)$ is, instead, likely due to spin waves of the non-collinear AIAO magnetic state of Ir spins.

Polycrystalline samples of Y$_{2-x}$Bi$_x$Ir$_2$O$_7$ were grown via a multistep process as described in the supplemental material. The salient aspect of our procedure is that all steps were conducted in sealed tubes with excess IrO$_2$ and long reaction times. The pyrochlore structure is generally known to support two types of defects causing non-stoichiometry: A-site vacancy and anti-site mixing for the A and B sites in A$_2$B$_2$O$_7$. As noted recently, in situations where x-ray diffraction can cleanly distinguish between these two scenarios, all evidence points to an A-site vacancy scenario [16]. Thus we propose that the sealed environment prevents the loss of Y and the excess Ir provides sufficient partial pressure to counteract Ir loss, both measures promoting the formation of stoichiometric material. Measurements of $\rho(T)$ were made in a Physical Property Measurement System (PPMS) using the standard four-probe method over the temperature range 10-300K. Due to the large magnitude of $\rho(T)$ in the 3-10K range for x = 0, $\rho(T)$ was measured using the 100MΩ range of an Agilent 34401A multimeter using the known internal 10 MΩ resistance in parallel with the unknown input resistance while applying a 500 nA current. Measurements of $C(T)$ were made using the relaxation method in the PPMS.

In Fig. 1, we present $\rho(T)$ for our sample, and we see $\rho(T) \propto T^{-4}$ from 3K to 50K, thus over more than four orders of magnitude in $\rho$. We attempted to fit the data to forms commonly used for modeling $d\rho/dT < 0$ behavior: $\rho_q(T) = \rho_0 \exp(\Delta/T^q)$ where $q$ = 0.25, 0.5, and 1 for variable range hopping, variable range hopping with electron correlations, and either intrinsic semiconducting or nearest neighbor hopping, respectively. In Fig. 1 we show the results of least-squares fits of $\ln(\rho)$ vs. $\ln(T)$ to $\rho_q$ models for q = 0.25, 0.5, and 1 as well as a fit to the TSCI WSM model [4] $\rho = \rho_0 T^{-4}$, which gives the best fit. We note that an alternative model for $\rho(T)$ in an undoped, pure, and interacting electron WSM yields $\rho(T) \propto T^{-1}$. We believe the inapplicability of this model highlights the importance of scattering from charged impurities.

Previously published $\rho(T)$ data for Y$_2$Ir$_2$O$_7$ are highly irreproducible. For instance, $\rho(10K)$ spans four orders of magnitude among different samples and, while $\rho(T)$ generally exhibits $d\rho/dT < 0$, no common functional form is found [11,15,17-21]. This variability can be attributed to a sensitivity of iridate pyrochlores to doping-type defects and, for Y$_2$Ir$_2$O$_7$, is likely related to A-site vacancy formation or oxygen defects [16]. It is important to note that, while the variability among previously studied samples is large, to the extent that a power law $\rho(T) \propto T^{-\eta}$ can be identified, $\eta \leq 4$ has always been observed [11,15,17,19-23]. In addition, we know of no case where modification of an iridate pyrochlore, either by oxidizing or by substitution, leads to insulating behavior and, as shown in [16], single crystals exhibit the largest magnitudes of $\rho(T)$. Hence a grain boundary origin of $T^{-4}$ behavior is unlikely. In other words, our sample exhibits behavior that is *i*) the limiting behavior among existing measurements and *ii*) the expected behavior for an undoped WSM.

In order to test the stability of the exponent $\eta = 4$ in Y$_2$Ir$_2$O$_7$, we substituted Bi for Y at concentrations lower than in previous studies. In Fig. 2 are shown $\rho(T)$ data for Y$_{2-x}$Bi$_x$Ir$_2$O$_7$ with x = 0.02, 0.04, and 0.1. Previous work had demonstrated the onset of metallic behavior, $d\rho/dT > 0$, for x = 0.3 above 120K, below which charge localization leads to $d\rho/dT < 0$ [21]. At our lower Bi concentrations, we observe $d\rho/dT < 0$ for all of our samples, and find exponents $\eta = 4.1(1), 3.9(1), 1.7(1)$ for x = 0.02, 0.04, and 0.1 respectively. Thus, $\eta = 4$ behavior is found to be stable for Bi concentrations up to x = 0.04, i.e. 2% on the Y site. Clearly, Bi incorporation does not represent doping in the usual sense of charge transfer from the substituted atom to the electronically-active bands since Bi and Y are isovalent at 3+. It is also unlikely that this metal-insulator crossover is due to bandwidth control since the ionic radius of Bi$^{3+}$(1.11 Å) is greater than that of Y$^{3+}$(1.015 Å), and the lattice constant, *a*, for Y$_{2-x}$Bi$_x$Ir$_2$O$_7$ obeys a Vegard's law for x = 0 – 2: $a = 10.167 + 0.0765x$ [22]. This behavior, coupled with the observed reduction in $\rho(T)$ with x implies $d\rho/dP > 0$, which is counter to that seen in all studies of R$_2$Ir$_2$O$_7$ under pressure, where $d\rho/dP < 0$ [24]. Thus a different mechanism is needed to explain the metal-insulator crossover in Y$_{2-x}$Bi$_x$Ir$_2$O$_7$ and we will return to the effect of Bi substitution later when discussing the $C(T)$ and the magnetism of the Ir sublattice.

The stability of the $\rho(T) = \rho_0 T^{-4}$ behavior, the prediction of WSM behavior in Y$_2$Ir$_2$O$_7$, and the lack of an alternative explanation strongly suggests the applicability of the TSCI model to our data [6]. The coefficient $\rho_{0,Weyl}$ in this model is given by [5]

$$\rho_{0,Weyl} = \frac{180\alpha^2 |\ln(g\alpha)| n_{imp} \hbar^5 v^4}{7g\pi^2 e^2 k_B^4} \qquad \text{Eq. 1}$$

where $\alpha$ is the fine structure constant, $g$ the Weyl node and spin degeneracy, $n_{imp}$ the charged impurity density, and $v$ the velocity. It is worth noting that power law dependence of $\rho(T)$ is quite unusual. In the present case, one can think of $T^{-4}$ as arising from the product of a term related to the density of states $g(E) \propto E^2$, which contributes two powers of $T$, and a scattering cross section which contributes two more powers of $T$. To compare with our measured $\rho_0 = 2.5 \times 10^9 \, \Omega \, \text{cm} \, K^4$, we use $v = 10^8$ cm/s [25], $g = 48$ [8], and assume $n_{imp} = 10^{18}$ ($10^{19}, 10^{20}$) cm$^{-3}$, and $\alpha = e^2/4\pi\epsilon\epsilon_0 v\hbar = 2.2$ (0.79, 0.29) for a dielectric constant $\epsilon = 1$ (2.8, 7.6). Substituting these values into Eq. 1 yields $\rho_{0,Weyl} = 1.7 \times 10^9 \Omega \text{cm} K^4$, close to our measured value. Because the materials parameters used are of reasonable magnitudes, the similarity between the measured and predicted values of $\rho_0$ lends further credence to associating our $\rho(T) = \rho_0 T^{-4}$ with behavior of an undoped WSM. It is natural, however, to ask if $\rho(T)$ is actually probing a combination of bulk and surface channels due to the polycrystalline nature of our samples. For this, we consider the magnetoresistance (MR). The inset of Fig. 1 shows that the MR is very small – even at the low temperature of 5K, MR = 3% at 8T – which is in contrast to much larger MRs exhibited by doped WSM such as TaAs [27]. This small MR in Y$_2$Ir$_2$O$_7$ can be understood however, by calculating the magnetic length for cyclotron motion, $l_m = 2v/\omega_c$ where $\omega_c = ev^2H/2k_bT$ is the cyclotron frequency for a Weyl electron. The ratio of magnetic length to mean free path is thus $l_m/l = (4ge/3\pi^2)(\rho/H)\lambda_D^{-3}$, where $\lambda_D = \hbar v/k_B T$ is the deBroglie wavelength. Using the measured value of $\rho$ at 5K yields $l_m/l = 1.5 \times 10^3$ at $H = 8$T, implying a negligible effect of field on electron motion. In addition, the dominant contribution to MR at 5K is negative, whereas a positive MR is expected from orbital motion. The most likely cause of MR < 0 is a field-induced reduction of spin-wave scattering. Thus, the MR data are consistent with bulk transport.

Charge transport in metals is often parameterized by the Ioffe-Regel parameter $k_F l$, where $k_F$ is the Fermi wave-vector, $l$ is the mean free path, and $k_F l > 2\pi$ corresponds to $l$ greater than a lattice constant. The analogous parameter for an undoped WSM is $k_T l$, where $k_T = k_B T/\hbar v$ is the wave-vector of thermally generated fermions. We can thus obtain an estimate of $l$ from the Drude resistivity, $\rho^{-1} = (ge^2/3\pi^2 \hbar)k_T^2 l$ [26] using our measured values of $\rho$. We find that $k_T l$ is $6.5 \times 10^{-9}$ and $1.6 \times 10^{-4}$ at 3K and 60K respectively. Thus, the non-interacting quasiparticle picture of transport breaks down in the region where $\rho \propto T^{-4}$ is observed. This breakdown is not unanticipated by the TSCI model where $\rho \propto T^{-4}$ is valid only below a temperature defined as $T_{imp} \approx (n_{imp} v^3/g)^{1/3} \hbar/k_B$, the cross over temperature between interaction dominated transport and impurity dominated transport, and above a temperature defined as $T_{min} \approx g^{-1/6} \alpha^{1/2} n_{imp}^{1/3} v\hbar/k_B$, below which a maximum resistivity associated with charge puddle formation is expected [5]. For the $Y_2Ir_2O_7$ materials parameters used above, $T_{imp} \sim 210K$, whereas $T_{min} \sim 590\ K$. Since $T < T_{imp}$ is consistent with the range of $\rho \propto T^{-4}$ behavior, we suggest that $T_{min}$ is not accurately predicted by the TSCI model and, given its functional form, cannot be brought into agreement with experiment by manipulation of the materials parameters. Given the above physical interpretation of $T_{min}$, however, it seems likely that inclusion of strong correlations, which should inhibit puddle formation, may provide an explanation for the temperature range over which $\rho(T) \propto T^{-4}$. Such correlations are motivated by Wan et al. [8] whose global LSDA + U + SO phase diagram predicts a WSM phase for 1.0 < U < 1.6eV. The prospect that the observed $\rho(T)$ is strongly influenced by electron correlations is reminiscent of so-called "bad metal" behavior as observed in some strongly correlated *metallic* systems, with $SrRuO_3$ [28] being a prime example. The hallmark of such systems is $k_F l < 2\pi$ as $T$ increases, depicting an unphysical picture of scattering at a length scale less than the primitive lattice spacing. In this regard, we note that our Drude-derived $l$ value at 60K, the upper limit of our $T^{-4}$ region, is $l \cong 0.2$Å and decreases further with decreasing temperature – thus the entire $T^{-4}$ region could be called, by analogy, a region of "bad" WSM behavior. Clearly further theoretical work is needed to explore the origins of such behavior.

Finally, we address $C(T)$ for pure and Bi-doped compounds. Previous work showed a qualitative difference in $C(T)$ between $x = 0$, which displays a rounded hump in $C/T$ vs. $T^2$, and $x \geq 0.1$ compounds, which show no hump and an intercept evoking a Sommerfeld $\gamma$-coefficient consistent with the appearance of metallicity. In Fig. 3 is presented $C/T$ vs. $T^2$ for x = 0, 0.04 and 0.10, which shows the evolution of the rounded hump in the low-Bi region. It is tempting to interpret the increase in slope below $T^2 = 80K^2$ for $x = 0$ as an indicator of the $C(T) \propto T^3$ behavior expected for Dirac dispersion [29]. Indeed, such an explanation has been advanced to explain a similar-looking hump in Ce$_3$Bi$_4$Pd$_3$ [30]. In the case of Y$_2$Ir$_2$O$_7$, however, since the WSM contribution to $C(T)$ is given by $C = \frac{56}{5}\pi^2 k_B (\frac{k_B T}{\hbar v})^3$ (calculated using a linear dispersion and Fermi distribution function), and with the above value of $v$, one expects a term given by the solid green line in Fig. 3. Such a contribution is four orders of magnitude less than the observed $C(T)$, making a WSM interpretation unlikely. An explanation for this hump in terms of phonons is also unlikely, given the very low Bi concentrations, which imply a negligible effective mass difference. A possible explanation for this feature is found, however, in the magnon contribution for non-collinear antiferromagnets [31]. As argued theoretically by Wan et al. and later determined by Disseler using a probability distribution analysis of µSR data [32], Y$_2$Ir$_2$O$_7$ possesses AIAO order, the non-collinearity of which leads to a spin wave dispersion that is not a simple function of the wavevector. The $C(T)$ for such AF order has been calculated and bears some similarity to the observed $C(T)$ on the low-temperature side of the hump, as shown in Fig. 3 inset. The association of the hump with magnons is also supported by the lack of field dependence also shown in Fig. 3. This field dependence has two implications. First, it demonstrates that the hump cannot be due to Schottky impurities, as discussed by Kimchi et al. [33]. Second, if the hump has a magnetic origin, the lack of field dependence at 8T implies a mean field energy scale much greater than 10K, consistent with $T_{Néel} = 150K$ for the Ir-moments. Normally, one would not expect the magnon contribution to be affected by a few percent substitution on the non-magnetic site. In the present case, however, such low substitution levels effect changes in $\rho(T)$ and lead to metallic behavior even at 15% Bi substitution. As Wan et al. argue, however, collinearity of the Ir spins is tightly correlated to metallic behavior hence it is reasonable to interpret the vanishing of the hump in $C/T$ vs. $T^2$ as evidence for the transition from excitation of a non-collinear state to a collinear state induced by the onset of metallicity. Such an interpretation is supported by the high-temperature susceptibility, $\chi(T)$. While a parasitic ferromagnetic signal has been observed at the Ir-moment-$T_{Néel}$ in most of the R$_2$Ir$_2$O$_7$ systems, the susceptibility above $T_{Néel}$ will contain information about the mean-field energy. As we see in Fig. 3 inset, the Weiss constant, $\theta_W$ demonstrates a dramatic jump between x = 0 and x = 0.02. This jump implies that the magnon $C(T)$ contribution will correspondingly decrease between x = 0 and 0.02, qualitatively consistent with observations. While more work is necessary to understand the precise interplay of low energy magnetic excitations and electronic transport, it appears as though the AIAO structure thought to exist for x = 0 is actually destabilized by small concentrations of Bi. Neutron scattering measurements are needed to identify the ordered state for small Bi concentrations.

In summary we have presented detailed measurements of the electrical resistivity and specific heat at low Bi concentrations of $Y_{2-x}Bi_xIr_2O_7$ for samples grown in a closed environment. These measurements showcase the robustness of the $\rho \propto T^{-4}$ power law and suggest a "bad" WSM scenario. In addition, a small observed MR is consistent with a small ratio of mean free path to magnetic scattering length. This small MR, combined with the power law $1/T^4$ makes $Y_2Ir_2O_7$ an attractive material for low-temperature thermometry. The low temperature $C(T)$ is understood in terms of magnon excitation of the AIAO structure, which is rapidly destabilized with Bi incorporation. Further theoretical and experimental work is needed to explore the ramification of strong electron correlations in the $R_2Ir_2O_7$ WSM's.

**Acknowledgments:** This work was supported by NSF-DMR 1534741 (P.G.L., A.P.R.), NSF DGE-1339067 (J.T.), and NSF-DMR 1534818 (T.S., L.D.). We would like to acknowledge useful conversations with S. Syzranov and B. Skinner.

References:

[1] S. M. Huang *et al.*, Nat Commun **6**, 7373 (2015).
[2] Q. Li *et al.*, Nature Physics **12**, 550 (2016).
[3] T. Besara *et al.*, Physical Review B **93**, 245152 (2016).
[4] Y. I. Rodionov and S. V. Syzranov, Physical Review B **91**, 195107 (2015).
[5] Y. I. Rodionov and S. Syzranov, arXiv:1503.02078 [cond-mat.mes-hall] **91**, 195107 (2015).
[6] S. Das Sarma, E. H. Hwang, and H. Min, Physical Review B **91**, 035201 (2015).
[7] D. Pesin and L. Balents, Nature Physics **6**, 376 (2010).
[8] X. Wan, A. M. Turner, A. Vishwanath, and S. Y. Savrasov, Physical Review B **83**, 205101 (2011).
[9] M. J. Graf *et al.*, Journal of Physics: Conference Series **551**, 012020 (2014).
[10] J. J. Ishikawa, E. C. T. O'Farrell, and S. Nakatsuji, Physical Review B **85**, 245109 (2012).
[11] H. Kumar, R. S. Dhaka, and A. K. Pramanik, Physical Review B **95**, 054415 (2017).
[12] K. Matsuhira, M. Wakeshima, Y. Hinatsu, and S. Takagi, Journal of the Physical Society of Japan **80**, 094701 (2011).
[13] W. Witczak-Krempa, G. Chen, Y. B. Kim, and L. Balents, Annual Review of Condensed Matter Physics **5**, 57 (2014).
[14] W. Witczak-Krempa, A. Go, and Y. B. Kim, Physical Review B **87**, 155101 (2013).
[15] D. Yanagishima and Y. Maeno, Journal of the Physical Society of Japan **70**, 2880 (2001).
[16] A. W. Sleight and A. P. Ramirez, Solid State Communications **275**, 12 (2018).

[17] H. Fukazawa and Y. Maeno, Journal of the Physical Society of Japan **71**, 2578 (2002).
[18] H. Kumar and A. K. Pramanik, Journal of Physics: Conference Series **828**, 012009 (2017).
[19] H. Kumar and A. K. Pramanik, Journal of Magnetism and Magnetic Materials **409**, 20 (2016).
[20] H. Liu, W. Tong, L. Ling, S. Zhang, R. Zhang, L. Zhang, L. Pi, C. Zhang, and Y. Zhang, Solid State Communications **179**, 1 (2014).
[21] M. Soda, Physica B: Condensed Matter **329-333**, 1071 (2003).
[22] N. Aito, M. Soda, Y. Kobayashi, and M. Sato, Journal of the Physical Society of Japan **72**, 1226 (2003).
[23] S. M. Disseler, C. Dhital, A. Amato, S. R. Giblin, C. de la Cruz, S. D. Wilson, and M. J. Graf, Physical Review B **86**, 014428 (2012).
[24] F. F. Tafti, J. J. Ishikawa, A. McCollam, S. Nakatsuji, and S. R. Julian, Physical Review B **85**, 205104 (2012).
[25] F. Ishii, Y. P. Mizuta, T. Kato, T. Ozaki, H. Weng, and S. Onoda, Journal of the Physical Society of Japan **84**, 073703 (2015).
[26] B. I. Shklovskii and A. L. Efros, *Electronic properties of doped semiconductors* (Springer Science & Business Media, 2013), Vol. 45.
[27] X. Huang *et al.*, **5**, 031023 (2015).
[28] L. Klein, J. S. Dodge, C. H. Ahn, G. J. Snyder, T. H. Geballe, M. R. Beasley, and A. Kapitulnik, Physical Review Letters **77**, 2774 (1996).
[29] H. H. Lai, S. E. Grefe, S. Paschen, and Q. M. Si, Proceedings of the National Academy of Sciences of the United States of America **115**, 93 (2018).
[30] S. Dzsaber, L. Prochaska, A. Sidorenko, G. Eguchi, R. Svagera, M. Waas, A. Prokofiev, Q. Si, and S. Paschen, Phys Rev Lett **118**, 246601 (2017).
[31] N. Arakawa, Journal of the Physical Society of Japan **86**, 094705 (2017).
[32] S. M. Disseler, Physical Review B **89**, 140413 (2014).
[33] I. Kimchi, J. P. Sheckelton, T. M. McQueen, and P. A. J. a. p. a. Lee, (2018).

Figures:

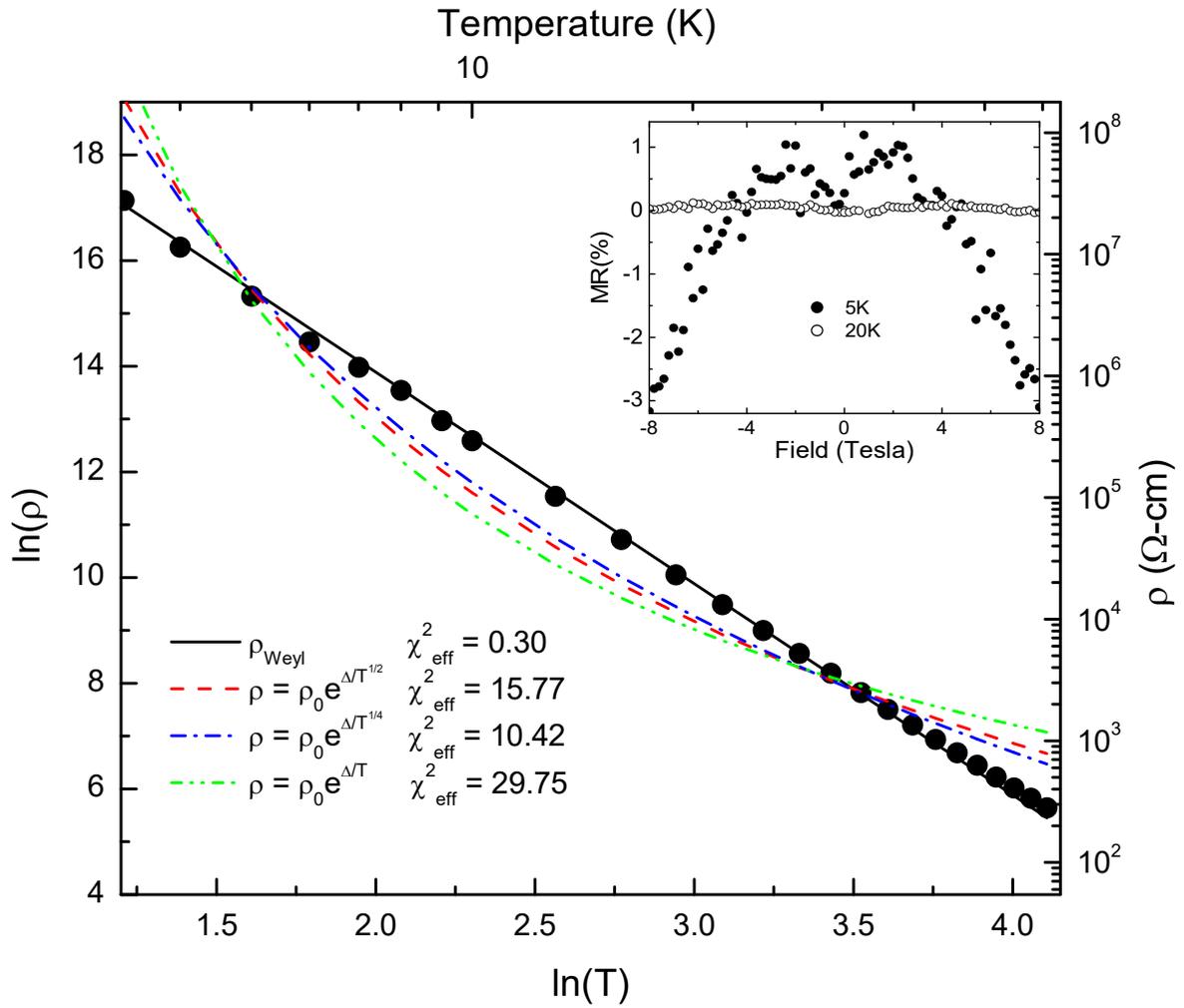

**Figure 1.** Resistivity versus temperature of $Y_2Ir_2O_7$. The solid black line is the theoretical dc resistivity of a WSM with compensated charge impurities. The solid blue, red, and green lines are best fits of the resistivity for variable range hopping and hopping with electron-electron interactions, and intrinsic semiconducting behavior with $\rho_0 = 3.3 * 10^{-4}, 8.9 * 10^{-9}$, and 6.10 $\Omega$cm, and $\Delta = 29.8$ K$^{\frac{1}{2}}$, 48.7 K$^{\frac{1}{4}}$, and 21.6K respectively.

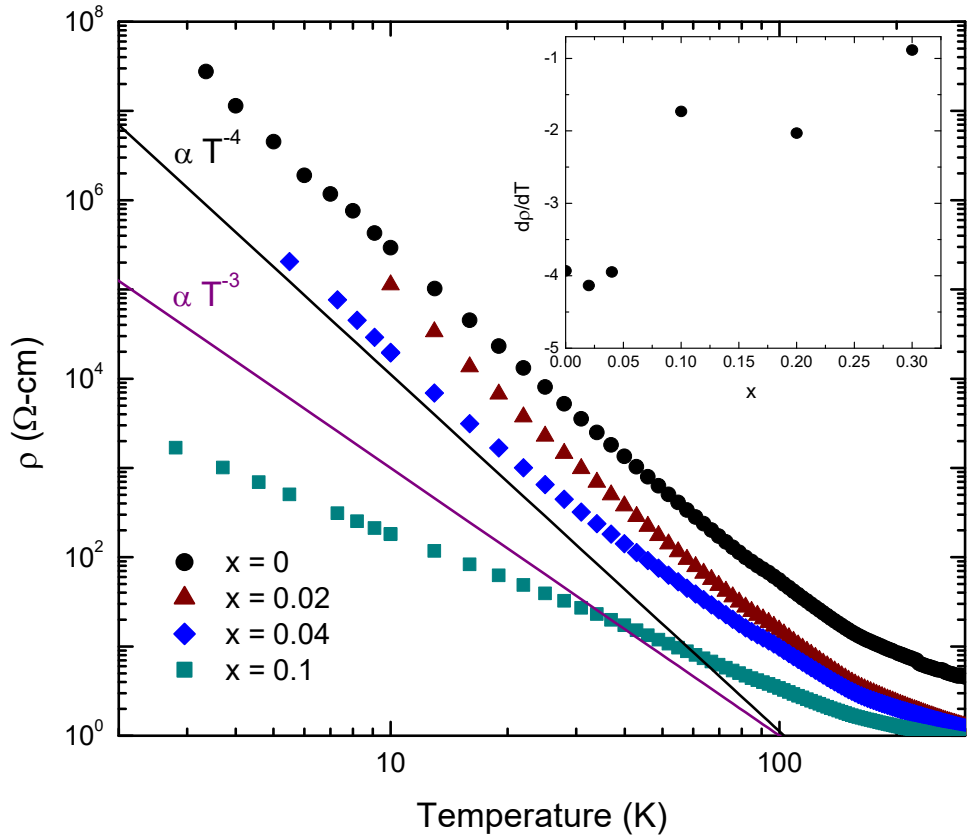

**Figure 2.** Resistivity versus temperature of $Y_{2-x}Bi_xIr_2O_7$. The inset is the derivative of resistivity with respect to temperature for different concentrations. For concentrations greater than x = 0.04 transport deviates from $T^{-4}$.

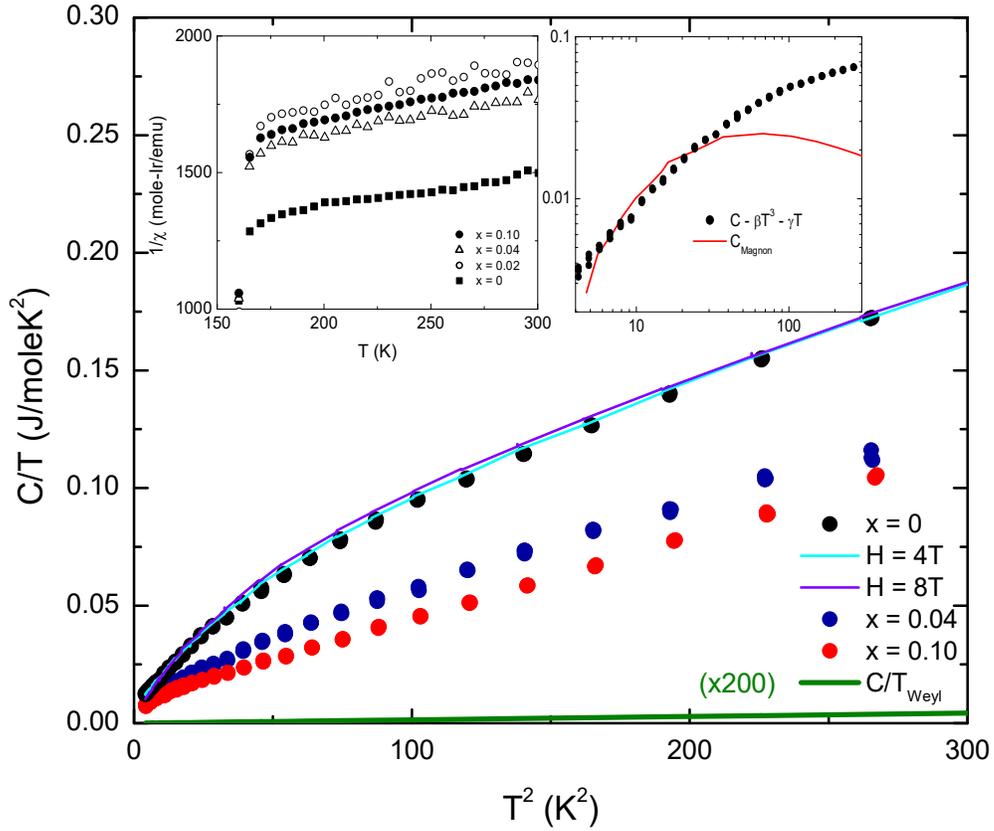

**Figure 3.** The specific heat divided by temperature for $Y_{2-x}Bi_xIr_2O_7$, as a function of temperature squared for specific concentrations of Bi doping. The right inset shows the comparison of the specific heat for the $x = 0$ sample without the electronic contribution to that of the theoretical specific heat contribution due to magnons of a magnetic pyrochlore system with AIAO ordering. The left inset shows the inverse susceptibility as a function of temperature for different concentrations of Bi.

Supplemental Material

Evidence for Undoped Weyl Semimetal Charge Transport in $Y_2Ir_2O_7$

P.G. LaBarre et al.

**Synthesis of $Y_{2-x}Bi_xIr_2O_7$**

Polycrystalline samples of $Y_2Ir_2O_7$ were obtained via a multistep process in closed silica ampoules. First, stoichiometric amounts of $Y_2O_3$ and $IrO_2$ were thoroughly mixed and pressed into a pellet. The pellet was placed in an alumina crucible and sealed under vacuum in a silica ampoule. The ampoule was placed in a furnace and heated to 1000 ºC for 48 hours. This procedure was repeated until most of the starting material was reacted. Due to a sluggish solid state reaction, however, no phase-pure material was obtained in this way. The next step includes the addition of 10% excess $IrO_2$, a top temperature of 1150 ºC that is slightly higher than the melting point of $IrO_2$, and an atmosphere of about 110 MPa of argon. In this way, a substantial $IrO_2$ partial pressure is established in the ampoule, and $IrO_2$ melting will ensure a more complete reaction. After 72 hours of heating at this temperature, the ampoule is cooled to 1000 ºC, and placed into a temperature gradient to transport excess $IrO_2$ away from the pellet. If needed, the process is repeated until powder X-ray diffraction indicates single-phase material [1].

Compared to polycrystalline $Y_2Ir_2O_7$, polycrystalline $Bi_2Ir_2O_7$ is easier to synthesize since $Bi_2O_3$ has a low melting point of 817 ºC and acts as a flux/mineralizer. Mixtures of $Bi_2O_3$ and Ir with proper molar ratios are ground, pressed into pellets, placed in alumina crucibles and heated at 1000 ºC for 48 hours, after which it is cooled to room temperature in the furnace. At this temperature, $IrO_2$ volatility is reduced sufficiently to maintain the stoichiometry. With only a few intermediate grindings and heat treatments, single-phase polycrystalline $Bi_2Ir_2O_7$ is obtained.

The synthesis method of bismuth alloyed yttrium iridates $Y_{2-x}Bi_xIr_2O_7$ proceeds similarly to $Bi_2Ir_2O_7$. Here, $Y_2O_3$, $Bi_2O_3$ and Ir with proper molar ratios were used as starting materials in case for a sufficiently large $Bi_2O_3$ content. The starting materials were ground, pressed into pellets, and heated at 1000 ºC for 24 hours, followed by cooling in the furnace. The processes of grinding and heat treatments were repeated until phase-purity was reached. However, for small x, not enough $Bi_2O_3$ is present to ensure rapid uniform phase formation. To overcome this limitation and synthesize $Y_{2-x}Bi_xIr_2O_7$ with the desired bismuth content, $Y_2Ir_2O_7$ and $Bi_2Ir_2O_7$ were used as starting materials. Mixtures of $Y_2Ir_2O_7$ and $Bi_2Ir_2O_7$ with proper molar ratios were ground, pressed into pellets, placed in alumina crucibles and heated at 1000 °C for 48 hours after which they are cooled to room temperature in the furnace. To ensure that the reaction is complete, grinding and heat treatments are repeated several times [1].

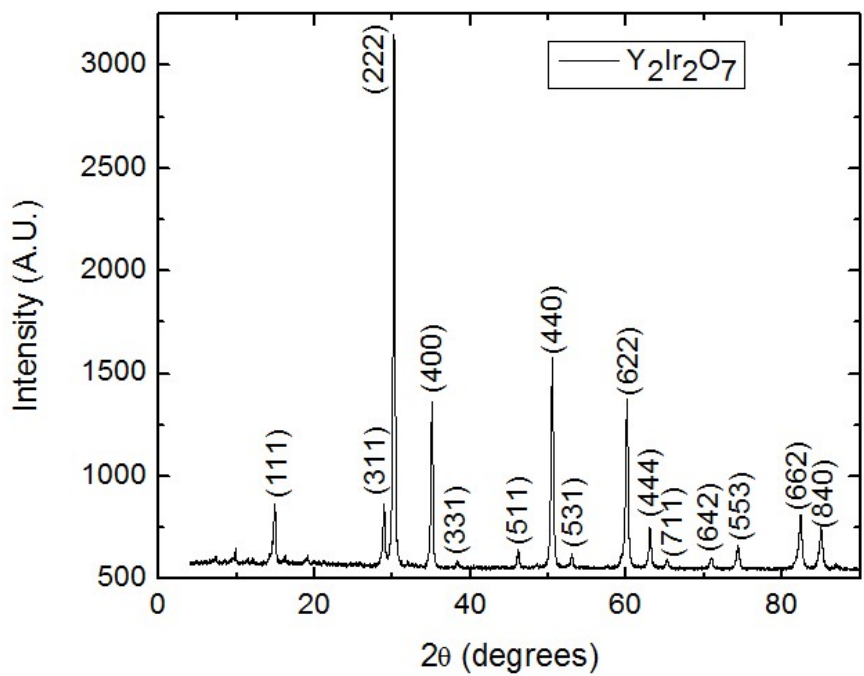

Fig. S1 X-ray powder pattern (Cu K$\alpha$ radiation) of polycrystalline $Y_2Ir_2O_7$. All peaks are indexed to the pyrochlore structure.

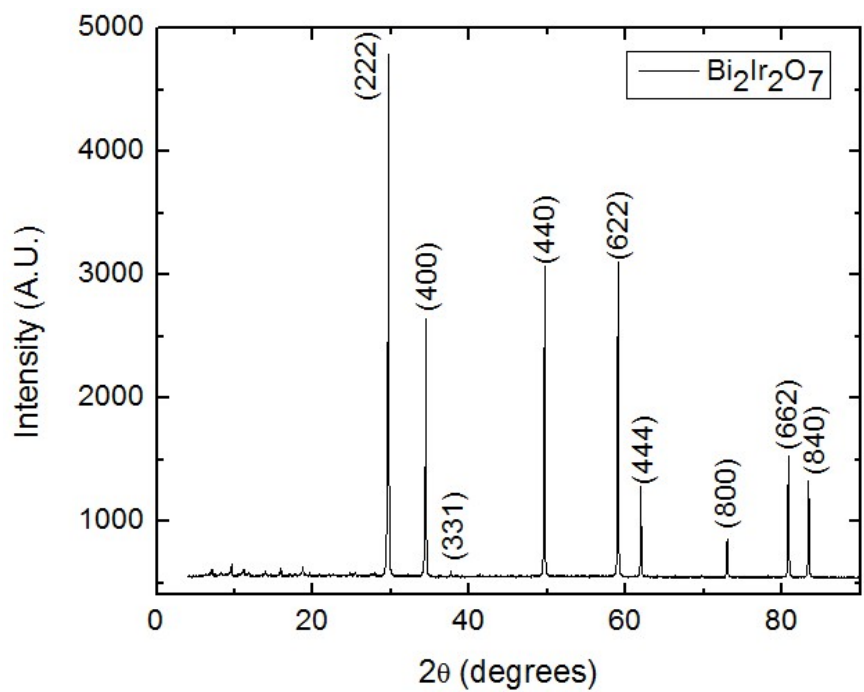

Fig. S2 X-ray powder pattern (Cu K$\alpha$ radiation) of polycrystalline $B_2Ir_2O_7$. All peaks are indexed to the pyrochlore structure.

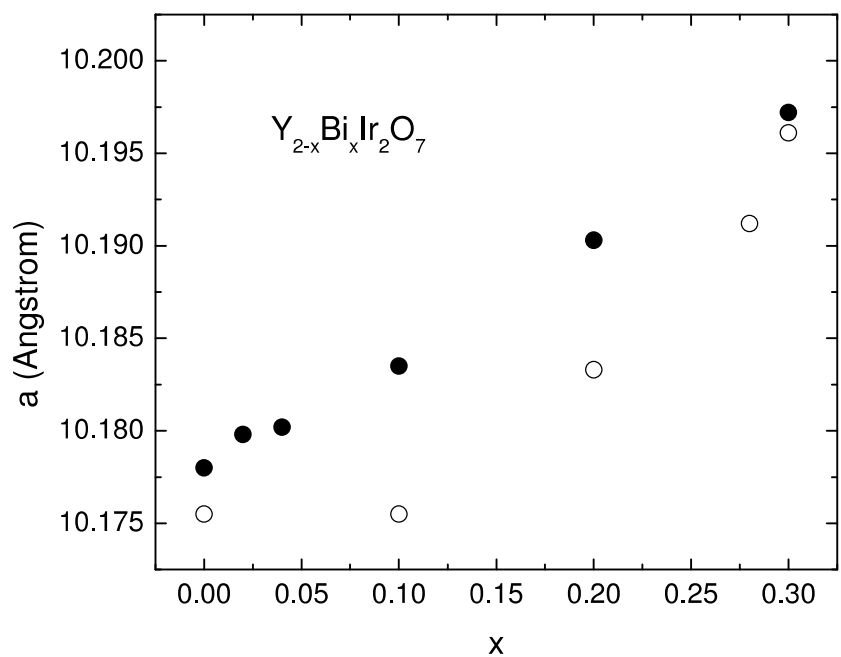

Fig. S3  Lattice parameters obtained from X-ray diffraction for the samples studied here (filled circles), compared to those obtained by Soda et al. [2]. The expected lattice expansion is observed to be systematic and proportional to Bi content in the low-Bi density region (x < 0.1) of importance for the present work.

1. Dong, L., "Ph.D. Thesis", *Florida State University*, (2017)
2. Soda, M., Aito, N., Kurahashi, Y., Kobayashi, Y., Sato, M., "Transport, thermal and magnetic properties of pyrochlore oxides $Y_{2-x}Bi_2Ir_2O_7$", *Physica B*, **329-333**, 1071 (2003).